\documentclass[twocolumn,english,aps,prl,nofootinbib,hidelinks]{revtex4-2}
\usepackage{lmodern}
\usepackage[latin9]{inputenc}
\usepackage{geometry}
\usepackage{color}
\usepackage{babel}
\usepackage{float}
\usepackage{amsmath}
\usepackage{amssymb}
\usepackage{graphicx}
\usepackage[unicode=true]
 {hyperref}

\makeatletter

\usepackage{orcidlink}
\usepackage{multirow}

\usepackage{txfonts}

\hypersetup{
    colorlinks=true,
    linkcolor=blue,
    citecolor=blue,
    urlcolor=blue
}
{\newpage}
\newcommand{\lb}{\linebreak}

\makeatother

\begin{document}
\title{Progenitor age-bias-corrected Type Ia supernovae favor a logarithmic
luminosity-distance relation}
\author{Hoang Ky Nguyen \orcidlink{0000-0003-2343-0508}$\ $}
\email[\ ]{hoang.nguyen@ubbcluj.ro}

\affiliation{{\vskip2pt}Department of Physics, Babe\c{s}-Bolyai University, Cluj-Napoca
400084, Romania\vskip1ptInternational Center for Interdisciplinary
Science and Education, ICISE, Quy Nhon 55121, Vietnam}
\date{\today}
\begin{abstract}
Type Ia supernovae provide a principal and direct observational probe
of late-time cosmic acceleration. Recently, Son et al. {[}\href{https://academic.oup.com/mnras/article/544/1/975/8281988}{MNRAS 544, 975 (2025)}{]}
proposed that correlations between SNe Ia luminosity and progenitor
age could introduce a systematic bias in the inferred luminosities,
leading to revised distance moduli in the Pantheon+ and DES-SN5YR
compilations. Motivated by this possibility, we test whether the progenitor
age-bias-corrected SNe Ia Hubble diagrams comply with the closed-form
luminosity-distance relation $d_{L}=\frac{c}{H_{0}}(1+z)\ln(1+z)$.
We find that this parsimonious logarithmic relation successfully describes
both age-bias-corrected datasets, with the Hubble constant $H_{0}$
as its sole free parameter for each individual compilation. For Pantheon+
and DES separately, the relation yields lower $\chi^{2}$ values than
the two-parameter flat $\Lambda$CDM baseline, with $\Delta\chi^{2}=12.0$
and $2.1$ respectively, and is favored by both the Akaike and Bayesian
information criteria. A joint likelihood analysis of both datasets
allowing independent $H_{0}^{\text{Pan+}}$ and $H_{0}^{\text{DES}}$
yields $\Delta\chi^{2}=17.1$, corresponding to $(\Delta\text{AIC},\Delta\text{BIC})=(19.1,25.2)$
in favor of the logarithmic relation. We further find that, after
the progenitor age-bias correction, neither dataset requires higher-order
corrections to the relation, whereas the Einstein--de Sitter cosmology
is firmly rejected. These results support the closed-form $d_{L}=\frac{c}{H_{0}}(1+z)\ln(1+z)$
relation as a viable one-parameter description of progenitor age-bias-corrected
SNe Ia Hubble diagrams. Its asymptotic behavior $d_{L}\propto z\,\ln z$
naturally accounts for the excess distance moduli observed at high
redshift. Should the progenitor age-bias correction be confirmed,
the logarithmic relation can be stringently tested using larger samples
of SNe Ia within $1\lesssim z\lesssim2$, where it already departs
sharply from flat $\Lambda$CDM.
\end{abstract}
\maketitle

\section{\label{sec:Introduction}Introduction}

Type Ia supernovae (SNe Ia) provide the primary observational evidence
for the late-time accelerated expansion of the Universe \citep{Riess,Perlmullter}.
Recently, Son et al. \citep{Son-2} reported evidence that correlations
between SNe Ia luminosity and host-galaxy age could introduce a systematic
bias in standardized SNe Ia luminosities, particularly\lb at high
redshift. The inferred bias reaches about $0.18$ mag ($5.5\sigma$)
at $z\sim1$. More recently, Zhang et al. \citep{Zhang} reported
a standardized-luminosity age step of 0.163 mag (5.2$\sigma$) between
SNe Ia in younger and older local environments. If confirmed, the
progenitor age-bias correction would reduce the inferred distance
moduli of high-redshift SNe Ia in the Pantheon+ and DES-SN5YR data.

Amid the ongoing debate over the validity of this correction \citep{Wiseman,Chung,Murakami,Popovic-2026},
analyses based on these revised Hubble diagrams have led to distinct
cosmological implications. Assuming the progenitor age-bias-corrected
Pantheon+ and DES-SN5YR Hubble diagrams, Son et al. \citep{Son-2}
and Sah et al. \citep{Sah} independently argued for the possibility
of a currently non-accelerating or decelerating universe. Mirpoorian
et al. \citep{Mirpoorian} investigated the cosmological impact of
a phenomenological redshift-dependent SNe Ia calibration correction,
without identifying it specifically with progenitor age. Redshift
dependence in the effective empirical calibration of SNe Ia has likewise
been investigated \citep{Mahtessian}.

In this Letter, we point out \emph{a third possibility}: the age-bias-corrected
data may instead be pointing toward a simple closed-form one-parameter
luminosity-distance relation
\begin{equation}
d_{L}=\frac{c}{H_{0}}(1+z)\ln(1+z).\vspace{-.1cm}\label{eq:dL-loga}
\end{equation}
We test this possibility through three complementary analyses: (i)
using Eq. \eqref{eq:dL-loga} to fit directly to the age-bias-corrected
Pantheon+ and DES Hubble diagrams, with flat $\Lambda$CDM serving
as a reference baseline; (ii) using the $w$CDM model as an intermediate
step to reach Eq. \eqref{eq:dL-loga}; and (iii) using a model-independent
kernel-construction method to infer Eq. \eqref{eq:dL-loga}.

We emphasize four key aspects of our approach:

\paragraph{1)}

Relation \eqref{eq:dL-loga} does not restore the Einstein--de Sitter
(EdS) cosmology, which is decisively rejected by the age-bias-corrected
Pantheon+ and DES data. The excess distance moduli at high redshift
remain incompatible with EdS.

\paragraph{2)}

Relation \eqref{eq:dL-loga} agrees with EdS (and flat $\Lambda$CDM)
at low redshift, recovering the Hubble law $d_{L}\simeq\frac{c}{H_{0}}z$.
Yet, despite involving only $H_{0}$ as the single free parameter,
its comoving distance increases logarithmically, $\text{\ensuremath{\frac{d_{L}}{1+z}}\ensuremath{\propto}}\ln(1+z)$,
rather than approaching the saturated EdS limit $\frac{d_{L}}{1+z}\propto2\,\bigl(1-\frac{1}{\sqrt{1+z}}\bigr)$.
Consequently, the relation naturally accounts for the excess distance
moduli observed at high redshift, a role traditionally attributed
to $\Omega_{\Lambda}$ in flat $\Lambda$CDM.

\paragraph{3)}

Its logarithmic form becomes markedly distinguishable from flat $\Lambda$CDM
for $z\gtrsim1$ (see Section \ref{sec: Kernel-bucket}). This opens
a feasible opportunity for a direct observational test through larger
samples of SNe Ia in $1\lesssim z\lesssim2$, a redshift regime already
accessible with current facilities.

\paragraph{4)}

Despite its compact closed form and the absence of any parameter representing
dark energy, the logarithmic relation \eqref{eq:dL-loga} can be derived
from several cosmological models, including Refs. \citep{Kolb,Melia,Nguyen-JCAP,Nguyen-EPJC}.
As these models embody underlying physics beyond standard $\Lambda$CDM
cosmology, they are expected to exhibit distinct phenomenology for
cosmological probes such as the baryon acoustic oscillations (BAO)
and the cosmic microwave background (CMB). One such model will be
discussed in Section \ref{sec:Discussion}.

To keep this Letter concise, methodologies are detailed in the Appendices,
and the Python scripts needed to reproduce the analysis are provided
at \textcolor{blue}{\small{}\url{https://github.com/HoangNguyenUBB/SNeIa-loga-relation}}.
The remainder of the Letter follows the three analyses outlined above
and concludes with a discussion.

\section{\label{sec:Direct-fit}Direct fit of the logarithmic relation to
age-bias-corrected SNe Ia Hubble diagrams}

Throughout this Letter, we refer to the original, uncorrected Pantheon+
and DES-SN5YR compilations as the pre-ABC data, and to the progenitor
age-bias-corrected compilations as the post-ABC data, where ABC denotes
the ``progenitor age-bias correction'' introduced by Son et al. \citep{Son-2}.
Following Son et al. \citep{Son-2}, we adjust the standardized SNe
Ia distance moduli of Pantheon+ and DES datasets by a \emph{redshift-dependent}
reduction $\Delta\mu$, where $\Delta\mu=0.183\,\Bigl(1-e^{-2.2z}\Bigr)$
is the analytic approximation to the correction curve shown in Fig.
2 of Ref. \citep{Son-2}. The correction equals 0.163 mag at $z=1$
and approaches 0.183 mag at high redshift.

We then test Eq. \eqref{eq:dL-loga} directly against the post-ABC
Pantheon+ and DES Hubble diagrams. Its only free parameter is the
Hubble constant $H_{0}$, fitted independently for each survey. Throughout
this work, the full covariance matrices supplied by the respective
collaborations are employed. The descriptions of data and conventions
used in this Letter are in Appendix \ref{app:Conventions}.

To exploit the statistical power of both Pantheon+ and DES surveys,
we conduct a joint likelihood analysis of the datasets. The joint
likelihood is provided in Appendix \ref{app:Joint-likelihood}. The
cosmological parameters $(\Omega_{\text{DE}},w)$ are common to both
surveys, whereas the Hubble constants $H_{0}^{\text{Pan+}}$ and $H_{0}^{\text{DES}}$
are fitted independently. This treatment allows each compilation to
have its own absolute distance modulus calibration while probing the
same underlying cosmology. The best-fitting parameters and goodness-of-fit
statistics are summarized in Table \ref{tab:best-fit-params}. For
Pantheon+ and DES separately, the logarithmic relation yields lower
minimum $\chi^{2}$ values than the two-parameter flat $\Lambda$CDM
model, despite involving only a single adjustable parameter. The corresponding
Akaike and Bayesian information criteria likewise favor the logarithmic
relation.

The result is noteworthy because Eq. \eqref{eq:dL-loga} contains
only a single free parameter, the present-day Hubble constant $H_{0}$,
while leaving its logarithmic shape unchanged. Consequently, the preferred
fit in Table \ref{tab:best-fit-params} reflects agreement with the
predicted shape itself rather than additional fitting freedom, such
as that provided by the shape parameter $\Omega_{\Lambda}$ in $\Lambda$CDM.
This conclusion is further strengthened by repeating the fits with
progressively higher maximum-redshift cutoff $z_{\text{max}}$, with
the preference first emerging at $z_{\text{max}}\simeq0.4$; see Table
\ref{tab:zmax} in Appendix \ref{app:Redshift-cutoff}. As a sanity
check, restricting the joint Pantheon+ and DES analysis to $z\leq1$
still yields $\Delta\chi^{2}=14.6$ in favor of the logarithmic relation,
indicating that the overall preference is not driven solely by the
highest-redshift SNe Ia (see Table \ref{tab:zmax}).
\begin{table}[!t]
\vspace{-.25cm}\setlength{\tabcolsep}{2pt}\caption{\label{tab:best-fit-params}Best-fit parameters and goodness of fit
for the logarithmic relation \eqref{eq:dL-loga} and flat $\Lambda$CDM
using post-ABC data. The joint analysis fits independent Hubble constants
for Pantheon+ and DES. Relative to flat $\Lambda$CDM, the logarithmic
relation lowers BIC by 19.4 (Pan+), 9.6 (DES), and 25.2 (Pan+ \& DES
jointly).\vspace{0.2cm}}

\small \begin{tabular}{cllccc} \hline\hline
\rule{0pt}{2.4ex}
Dataset & \ \ Model & \hspace{2em} Parameters & $\chi^2$ & AIC & BIC \\[0.4ex] \hline
\rule{0pt}{2.4ex}
\!Pan+ & Logarithmic & \ \ \hspace{0.3em}$H_0^{\rm Pan+}\!=72.56\pm0.12$ & 1749.5 & 1751.5 & 1756.9 \\[0.4ex]
Pan+ & $\Lambda$CDM & $\left\{\begin{array}{@{\hspace{0.3em}}r@{\;}c@{\;}l@{}} H_0^{\rm Pan+}\!& =&73.06\pm0.24\\ \Omega_\Lambda&=&0.462\pm0.023 \end{array} \right.$ & 1761.4 & 1765.4 & 1776.3 \\[2.4ex]
\hline
\rule{0pt}{2.4ex}
\!DES & Logarithmic & \ \ \hspace{0.3em}$H_0^{\rm DES}=68.79\pm0.15$ & 1635.5 & 1637.5 & 1643.0 \\[0.4ex]
DES & $\Lambda$CDM & $\left\{\begin{array}{@{\hspace{0.3em}}r@{\;}c@{\;}l@{}} H_0^{\rm DES}&=&70.16\pm0.36\\ \Omega_\Lambda&=&0.513\pm0.019 \end{array} \right.$ & 1637.5 & 1641.5 & 1652.6 \\[2.4ex]
\hline
\rule{0pt}{4ex}
\!Joint & Logarithmic & $\left\{\begin{array}{@{\hspace{0.3em}}r@{\;}c@{\;}l@{}} H_0^{\rm Pan+}\!\!&=&72.56\pm0.12\\ H_0^{\rm DES}&=&68.79\pm0.15 \end{array} \right.$ & $\bf{3384.9}$ & $\bf{3388.9}$ & $\bf{3401.3}$ \\[0.4ex]
Joint & $\Lambda$CDM & $\left\{\begin{array}{@{\hspace{0.3em}}r@{\;}c@{\;}l@{}} H_0^{\rm Pan+}\!\!&=&73.34\pm0.18\\ H_0^{\rm DES}&=&69.79\pm0.29\\ \Omega_\Lambda&=&0.491\pm0.014 \end{array}\right.$ & 3402.0 & 3408.0 & 3426.5 \\[3.6ex]
\hline\hline \end{tabular}\setlength{\tabcolsep}{3pt}\vspace{0cm}
\end{table}

\begin{figure}[!t]
\includegraphics[scale=0.57]{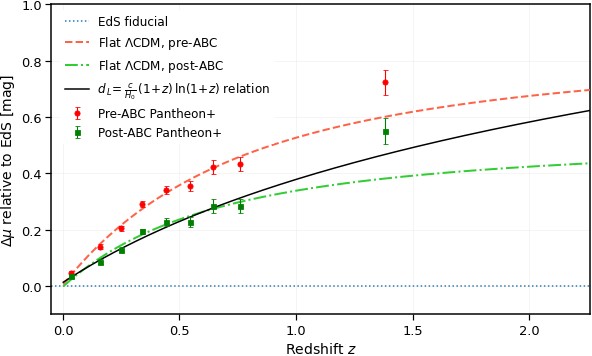}\vspace{-.2cm}

\caption{\label{fig:Direct-fit}Binned Pantheon+ Hubble diagram residuals relative
to a fiducial EdS model ($H_{0}=73$km/s/Mpc). The ABC substantially
reduces the high-redshift residuals but remains inconsistent with
the EdS prediction. The post-ABC data closely follow the logarithmic
luminosity-distance relation, whereas the pre-ABC data are better
described by the best-fitting flat $\Lambda$CDM model.}
\vspace{-.5cm}
\end{figure}

For visualization, Fig. \ref{fig:Direct-fit} compares the distance
modulus offsets of the best-fitting logarithmic relation, the flat
$\Lambda$CDM, and Pantheon+ with the EdS ($H_{0}=73$ fiducial) chosen
as benchmark. (The binning is carried out in Appendix \ref{app:Binning-Fig-1}.)
Evidently, the fiducial EdS is rejected irrespective of the inclusion
of the  ABC. Since the highest-redshift Pantheon+ bin $(1.0\leq z<2.3)$
contains only 25 SNe Ia, although these data favor the logarithmic
relation over flat $\Lambda$CDM, larger SNe Ia samples within this
redshift range are needed to test this apparent preference.

To test whether the post-ABC Hubble diagrams require additional flexibility
to fit, we further allow a quadratic correction to the logarithmic
kernel (see Appendix \ref{app:Mu2}). The additional quadratic term
is statistically consistent with zero for the joint datasets, indicating
that the post-ABC Hubble diagrams do not require higher-order modifications
to Eq. \eqref{eq:dL-loga}.

\emph{Despite its rigid shape and highly parsimonious form, the logarithmic
relation \eqref{eq:dL-loga} fits the post-ABC Pantheon+ and DES Hubble
diagrams remarkably well }\footnote{It is worth noting that, since the overall normalization $H_{0}$
can be supplied by local distance calibration, the logarithmic relation
\eqref{eq:dL-loga} is effectively a \emph{zero-parameter} description
of the SNe Ia Hubble diagram. The author thanks Junhyuk Son for drawing
attention to this interpretation.}\emph{, outperforming the (two-parameter) flat $\Lambda$CDM baseline
by $\Delta\text{BIC}=25.2$.} Furthermore, the relation \eqref{eq:dL-loga}
asymptotically behaves as $z\,\ln z$, naturally producing the excess
distance moduli observed at high redshift in SNe Ia Hubble diagrams
\emph{even though it contains no dark energy component}. The next
question is how it fares against the (three-parameter) flat $w$CDM
model.

\section{\label{sec:wCDM}The logarithmic relation in the flat $\boldsymbol{w}$CDM
parameter space}

The direct fit in Section \ref{sec:Direct-fit} demonstrates the empirical
viability of the parsimonious one-parameter logarithmic relation \eqref{eq:dL-loga}.
It is important to note that this relation corresponds to a special
case of the flat $w$CDM model, whose luminosity distance is\vspace{-.35cm}

\begin{equation}
\!\!\!d_{L}^{w\text{CDM}}\!=\!\frac{c}{H_{0}}(1+z)\!\!\int_{0}^{z}\!\!\!\frac{(1+z')\,d\ln(1+z')}{\sqrt{(1-\Omega_{\text{DE}})(1+z')^{3}+\Omega_{{\rm DE}}(1+z')^{3(1+w)}}}\label{eq:wcdm-dL}
\end{equation}
which reduces exactly to Eq. \eqref{eq:dL-loga} at $(\Omega_{{\rm DE}}^{*}=1,w^{*}=-1/3)$.
Within the flat $w$CDM parameter space, this special point corresponds
to the coasting universe, whose cosmic scale factor grows linearly
with cosmic time, $a\propto t$ \citep{Kolb}.

We then fit Eq. \eqref{eq:wcdm-dL} to the Pantheon+ and DES datasets,
using the joint likelihood described in Appendix \ref{app:Joint-likelihood}.
The cosmological parameters $(\Omega_{{\rm DE}},w)$ are shared by
both surveys, whereas the Hubble constants $H_{0}^{{\rm Pan+}}$ and
$H_{0}^{{\rm DES}}$ are fitted independently in order to account
for the independent absolute distance modulus calibrations of the
two compilations.

\begin{figure}[!t]
\includegraphics[scale=0.5]{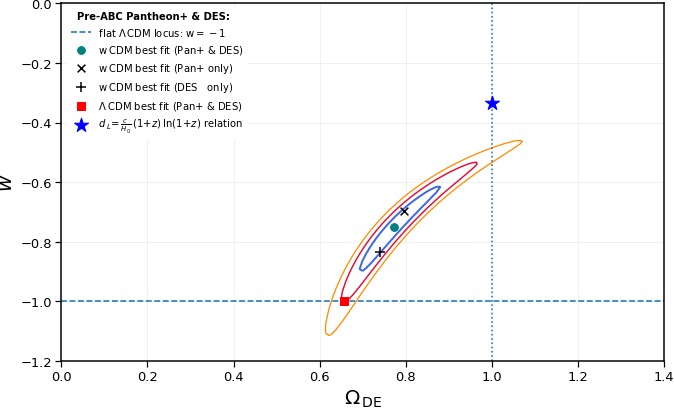}\vspace{0.2cm}

\includegraphics[scale=0.5]{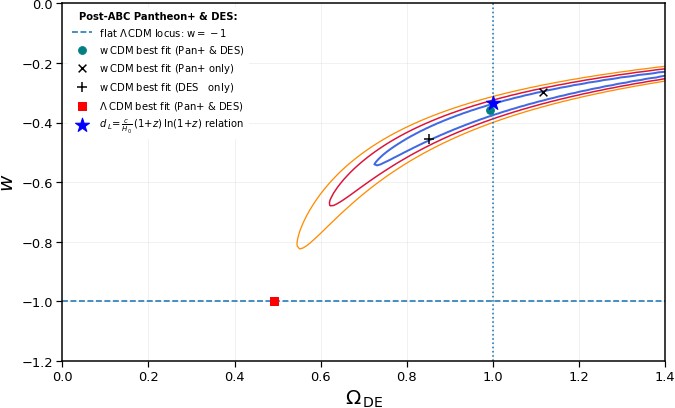}

\caption{\label{fig:Contours}Confidence contours (68.3\%, 95.5\%, and 99.7\%)
in the ($\Omega_{\text{DE}},w$) plane, obtained from the joint Pantheon+/DES
likelihood after profiling over $H_{0}^{\text{Pan+}}$ and $H_{0}^{\text{DES}}$.
Upper panel: pre-ABC Pantheon+/DES joint data. Lower panel: post-ABC
Pantheon+/DES joint data.}
\end{figure}

The resulting confidence contours are shown in Fig. \ref{fig:Contours}.
In contrast to the pre-ABC data (upper panel), the post-ABC data (lower
panel) exhibit two remarkable features:

\paragraph{\emph{(i)}}

The post-ABC degeneracy locus passes through the coasting point $(\Omega_{\text{DE}}^{*}=1,w^{*}=-1/3)$,
denoted by the blue star. Restricting the locus to $\Omega_{{\rm DE}}=1$
yields $-0.382\le w\le-0.331$ $(95.5\%\ {\rm CL})$, fully consistent
with the special value $w^{*}=-1/3$. The $w$CDM best-fit points
for individual Pantheon+ and DES datasets (cross and plus) also lie
along this special locus. We should stress that this alignment is
not \emph{a priori} expected---without the age-bias correction, the
locus does not run through the coasting point, as is evident in the
upper panel. \emph{The ABC therefore brings the coasting point }$(\Omega_{\text{DE}}^{*}=1,w^{*}=-1/3)$\emph{
into prominence.}

\paragraph{\emph{(ii)}}

Even more remarkably, in the lower panel, the $w$CDM best-fit point
for the post-ABC Pantheon+ and DES joint data nearly coincides with
the coasting point. The logarithmic relation, represented by the blue
star, is therefore achieved within the flat $w$CDM parameter space,
rather than being imposed \emph{a priori}. 

Do these two notable alignments imply that late-time cosmology obeys
the coasting universe? The answer is: \emph{Not necessarily.\lb}
Within the flat $w$CDM parameter space, the post-ABC SNe Ia data
favor the coasting point over the familiar flat $\Lambda$CDM by $\Delta\text{BIC}\simeq25$
(see Table \ref{tab:best-fit-params}). However, as will be discussed
in Section \ref{sec:Discussion}, \emph{the coasting point also belongs
to a broader family of cosmological models that share the logarithmic
relation \eqref{eq:dL-loga}}. This family intersects the flat $w$CDM
parameter space precisely at the coasting point. \emph{As such, the
coasting point marks a transition from the phenomenological flat $w$CDM
model to a broader family of cosmological models whose members can
embody richer physics and phenomenology than the coasting universe
itself.}

We next turn to a model-independent approach which allows us to extract
a latent structure of the post-ABC Hubble diagram.

\section{\label{sec: Kernel-bucket}Model-independent kernel construction}

The direct fit and the $w$CDM-based analysis in the two preceding
sections both adopt the logarithmic relation \eqref{eq:dL-loga} as
their starting point. Here we produce a third, entirely \emph{model-independent}
test of that relation. The key idea is that, from the \emph{observables}
$\{z,d_{L}\}$ of SNe Ia, we construct directly the dimensionless
quantity
\begin{equation}
K\equiv\frac{H_{0}}{c}\frac{d\left[d_{L}/(1+z)\right]}{d\ln(1+z)}\label{eq:Kernel-def}
\end{equation}
which we refer to as the \emph{luminosity-distance kernel} (or the
kernel for brevity), a procedure formulated in Appendix \ref{app:Kernel-forms}.
The kernel characterizes the \emph{observed} SNe Ia Hubble diagram
without assuming any cosmological model beforehand.

Theoretically, if the luminosity distance obeys the logarithmic relation
\eqref{eq:dL-loga}, then $K=1$ at every redshift. By contrast, the
kernel predicted by flat $\Lambda$CDM is a function of redshift $K_{\Lambda{\rm CDM}}=\frac{H_{0}}{aH_{\Lambda{\rm CDM}}}=\frac{1+z}{\sqrt{(1-\Omega_{\Lambda})(1+z)^{3}+\Omega_{\Lambda}}}$.
The two $d_{L}$ relations, i.e. Eq. \eqref{eq:dL-loga} versus flat
$\Lambda$CDM cosmology, therefore make two distinct predictions for
the kernel functional form.

To \emph{empirically} construct the kernel from the observables $\{z,d_{L}\}$,
we partition the combined Pantheon+ and DES sample into redshift buckets
as described in Appendices \ref{app:Kernel-forms} and \ref{app:Binning-Fig-3}.
Within each bucket, the kernel is obtained by a linear regression
of $d_{L}/(1+z)$ against $\ln(1+z)$. Interlaced buckets are also
used to verify that the construction is insensitive to the precise
placement of the bucket boundaries. The constructed values of $K$
are listed in Appendix \ref{app:Binning-Fig-3}, while the complete
implementation is provided at \textcolor{blue}{\small{}\url{https://github.com/HoangNguyenUBB/SNeIa-loga-relation}}.

\begin{figure}[!t]
\includegraphics[scale=0.57]{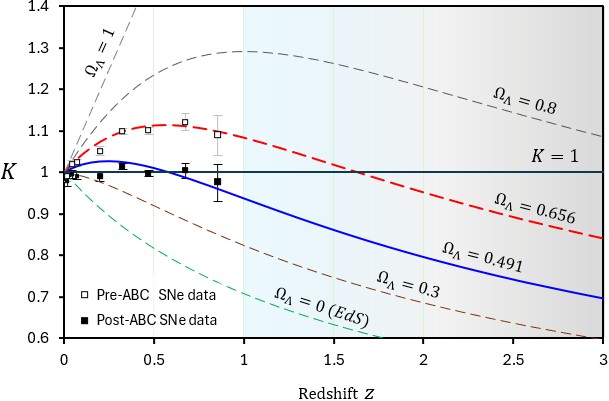}

\caption{\label{fig:K-bucket}Model-independent construction of the kernel
$K:=\frac{H_{0}}{c}\frac{d\left[d_{L}/(1+z)\right]}{d\ln(1+z)}$ from
the Pantheon+ and DES joint data. The kernel values (open squares)
obtained from the pre-ABC data closely match the prediction of flat
$\Lambda$CDM $(\Omega_{\Lambda}=0.656)$. The kernel values (solid
squares) are consistent with the logarithmic prediction $K=1$. By
contrast, the theoretical kernel of flat $\Lambda$CDM $(\Omega_{\Lambda}=0.491)$
departs systematically from the constructed kernel for $z\protect\geq1$
(shaded area).}
\vspace{-.2cm}
\end{figure}

Fig. \ref{fig:K-bucket} shows the constructed kernels for Pantheon+/DES
joint data. As a validation of our approach, the open squares, representing
the pre-ABC data, closely track the theoretical curve $K_{\Lambda\text{CDM}}(z;\Omega_{\Lambda}=0.656)$.

Upon applying the age-bias correction, three observations deserve
emphasis:

\paragraph{\emph{(i)}}

The solid squares, corresponding to the post-ABC data, are consistent
with the logarithmic prediction $K=1$ throughout the observed redshift
range. The preference for a constant kernel is already established
at low redshift. 

\paragraph{\emph{(ii)}}

Flat $\Lambda$CDM already shows a discrepancy from the $K$-bucket
construction. For the first four buckets $(0<z<0.5)$ listed in Table
\ref{tab:K-bucket-age-corrected}, the cumulative $\chi^{2}$ is 19.7
for flat $\Lambda$CDM kernel $K_{\Lambda\text{CDM}}(z;\Omega_{\Lambda}=0.491)$
and 13.5 for the logarithmic kernel $K=1$, giving $\Delta\chi^{2}=6.2$,
a value consistent with Table \ref{tab:zmax}. Thus, the discrepancy
is already manifest at low redshift rather than being driven by the
sparsely available highest-redshift SNe Ia. 

\paragraph{\emph{(iii)}}

The flat $\Lambda$CDM prediction departs systematically from the
constructed kernel for $z\gtrsim1$, again consistent with the highest\-z
bin (covering $z\ge1$) in Fig. \ref{fig:Direct-fit} of the direct
fit analysis. The interval $1\lesssim z\lesssim2$ is particularly
valuable because the logarithmic and flat $\Lambda$CDM predictions
start to diverge significantly there. Since this redshift range is
\emph{already} accessible to present and forthcoming SNe Ia surveys,
enlarging the SNe Ia sample over this interval provides the most direct
observational test capable of distinguishing Eq. \eqref{eq:dL-loga}
from flat $\Lambda$CDM.

Unlike the direct fit and the $w$CDM-based analysis, the kernel construction
makes no assumption about any underlying cosmological model. It therefore
independently verifies that the post-ABC Hubble diagram favors a nearly
constant logarithmic kernel. Taken together, assuming the ABC of Son
et al. \citep{Son-2}, the three analyses consistently support the
parsimonious one-parameter logarithmic relation \eqref{eq:dL-loga}.
\emph{Future SNe Ia observations, particularly over $1\lesssim z\lesssim2$,
offer the most direct opportunity to confirm or falsify the logarithmic
relation.}

\section{\label{sec:Discussion}Beyond late-time acceleration: \  How can
such a parsimonious luminosity-distance relation allow for the rich
phenomenology of cosmology?}

Among the cosmological models known to reproduce the logarithmic luminosity-distance
relation, Eq.\,\eqref{eq:dL-loga}, are the coasting cosmology of
Kolb \citep{Kolb} and Melia \citep{Melia}, together with a varying-speed-of-light
(VSL) realization presented in Ref.\,\citep{Nguyen-JCAP} constructed
from a scale-invariant theory of gravity and matter \citep{Nguyen-PLB,Nguyen-EPJC}.
More generally, it can be shown that the entire family of power-law
cosmologies satisfying $a(t)=\left(t/t_{0}\right)^{\,\mu}$, $c(a)=c_{0}\,a^{-\zeta}$,
with $(1+\zeta)\,\mu=1$, yields the same logarithmic relation \citep{Nguyen-2026}.
This follows immediately from the comoving distance along a radial
null geodesic in the flat FLRW metric,
\begin{equation}
r\,=\!\int_{t}^{t_{0}}\frac{c\,dt}{a}=\!\int_{t}^{t_{0}}\frac{c_{0}\,dt}{a^{1+\zeta}}=\frac{c_{0}t_{0}}{\mu}\!\int_{a}^{1}\!\!\frac{da}{a^{2+\zeta-\frac{1}{\mu}}}\,\propto\,\ln a\!\propto\!\ln(1+z)\label{eq:Dolgov-Barrow}
\end{equation}
due to $(1+\zeta)\,\mu=1$. Within this family, the VSL index $\zeta$,
which characterizes the power-law dependence of the speed of light
on the cosmic scale factor, may be regarded as a \emph{tunable} index,
analogous to the role phenomenologically played by $\Omega_{\Lambda}$
in flat $\Lambda$CDM. While the age-bias-corrected SNe Ia distance
moduli\lb comform to the logarithmic luminosity-distance relation
\eqref{eq:dL-loga}, they do not by themselves specify the \emph{latent}
VSL index. Since the speed of light governs many aspects of cosmology
beyond the late-time SNe Ia Hubble diagram---including BAO and the
CMB---these independent probes are expected to be sensitive to $\zeta$,
and can therefore, in principle, constrain its value.

Therefore, despite its \emph{deceptively} simple form and the absence
of any parameter representing dark energy, the logarithmic relation,
Eqs.\,\eqref{eq:dL-loga} and \eqref{eq:Dolgov-Barrow}, need not
be viewed merely as the coasting point $(\Omega_{\text{DE}}^{*}=1,w^{*}=-1/3)$
of the flat $w$CDM parameter space. Rather, it naturally encompasses
a broader class of VSL cosmological models,\emph{ whose phenomenological
richness is encoded in the VSL index} $\zeta$. This tunable index
provides the flexibility needed to confront other cosmological observations
beyond the late-time SNe Ia Hubble diagram, including BAO and the
CMB.

\vspace{.5cm}

\paragraph{Acknowledgments:}

HKN is deeply grateful to Junhyuk Son for generously providing the
Yonsei group's age-bias correction data, for his careful reading of
an early version of the manuscript, and for many insightful comments
and suggestions that substantially improved the presentation of this
work. He also thanks Tiberiu Harko and Soebur Razzaque for valuable
discussions.

\appendix
\onecolumngrid

\counterwithin{table}{section} \renewcommand{\thetable}{\Alph{section}.\arabic{table}}

\section{\label{app:Conventions}\ \  Data and Conventions for Reproducibility}
\begin{enumerate}
\item \emph{Input data:} The Pantheon+ and DES datasets are publicly available.
The analyses presented in this work use the following files and columns.
\begin{enumerate}
\item For Pantheon+ \citep{PantheonPlusData}:
\begin{itemize}
\item \small{\url{https://github.com/PantheonPlusSH0ES/DataRelease/tree/main/Pantheon%2B_Data/4_DISTANCES_AND_COVAR}}
\item Data file: Pantheon+SH0ES.dat; Columns used: zHD, zHEL, MU\_SH0ES
\item Covariance matrix: Pantheon+SH0ES\_STAT+SYS.cov
\end{itemize}
\item For DES-SN5YR \citep{DESData}:
\begin{itemize}
\item \small{\url{https://github.com/des-science/DES-SN5YR/tree/main/4_DISTANCES_COVMAT}}
\item Data file: DES-Dovekie\_HD.csv; Columns used: zHD, zHEL, MU
\item Covariance matrix: STAT+SYS.npz
\end{itemize}
\end{enumerate}
\item \emph{Redshift convention:} Throughout this work, the comoving distance
is evaluated using Hubble-diagram redshift $z_{\text{HD}}$, whereas
the luminosity-distance prefactor is evaluated using the heliocentric
redshift $z_{\text{hel}}$. This follows the conventions adopted in
the Pantheon+ and DES compilations. For example, Relation \eqref{eq:dL-loga}
is $d_{L}=\frac{c}{H_{0}}(1+z_{\text{hel}})\,\ln(1+z_{\text{HD}})$,
while the flat $\Lambda$CDM relation is $d_{L,\Lambda\text{CDM}}=\frac{c}{H_{0}}(1+z_{\text{hel}})\int_{0}^{z_{\text{HD}}}\frac{dz'}{\sqrt{(1-\Omega_{\Lambda})(1+z')^{3}+\Omega_{\Lambda}}}.$
\item \emph{Covariance matrix:} Unless stated otherwise, all parameter values
reported in this work are obtained using the full covariance matrices
of the pre-ABC and post-ABC Pantheon+ and DES datasets.
\item \emph{Definition of $\chi^{2}$:} The model parameters $\boldsymbol{\varTheta}$
are estimated by minimizing the statistic
\begin{equation}
\chi^{2}=\Delta\boldsymbol{\mu}(\boldsymbol{\varTheta})^{T}C^{-1}\Delta\boldsymbol{\mu}(\boldsymbol{\varTheta})
\end{equation}
where $\Delta\boldsymbol{\mu}(\boldsymbol{\varTheta})$ denotes the
vector of luminosity-distance modulus residuals, and $C$ is the covariance
matrix.
\item \emph{Scripts Repository:} The Python scripts needed to reproduce
the numerical analyses, tables, and figures presented in this work
are publicly available at\textbf{ }\small{\url{https://github.com/HoangNguyenUBB/SNeIa-loga-relation}}
\end{enumerate}

\section{\label{app:Joint-likelihood}\ \  Joint likelihood of Pantheon+
and DES surveys}

Assuming that the Pantheon+ and DES datasets are statistically independent,
their joint likelihood is
\begin{equation}
L_{\text{Joint}}(H_{0}^{\text{Pan+}},H_{0}^{\text{DES}};\boldsymbol{\varTheta})\ =\ L_{\text{Pan+}}(H_{0}^{\text{Pan+}};\boldsymbol{\varTheta})\ L_{\text{DES}}(H_{0}^{\text{DES}};\boldsymbol{\varTheta})\label{eq:joint-L-def}
\end{equation}
where $\boldsymbol{\varTheta}$ stands for a model's parameter set.
For the models considered in this Letter, the parameter set is
\begin{equation}
\boldsymbol{\varTheta}=\begin{cases}
\ \ \text{none} & \text{for the logarithmic relation \eqref{eq:dL-loga}, used in Section \ref{sec:Direct-fit};}\\
\ \ \text{none} & \text{for the luminosity-distance kernel \eqref{eq:Kernel-def}, used in Section \ref{sec: Kernel-bucket};}\\
\ \ \left\{ \,\delta\,\right\}  & \text{for the quadratic logarithmic relation \eqref{eq:Mu2-def}, used in Appendix \ref{app:Mu2};}\\
\ \ \left\{ \,\Omega_{\Lambda}\,\right\}  & \text{for \ensuremath{\Lambda}CDM model, used in Section \ref{sec:Direct-fit};}\\
\ \ \left\{ \,\Omega_{\text{DE}},\,w\,\right\}  & \text{for \ensuremath{w}CDM model, used in Section \ref{sec:wCDM}.}
\end{cases}
\end{equation}
Equation \eqref{eq:joint-L-def} reflects the fact that each survey
is characterized by its own nuisance parameter: the absolute distance
modulus $M^{\text{Pan+}}$ or $M^{\text{DES}}$. Equivalently, fitting
$M^{\text{Pan+}}$ and $M^{\text{DES}}$ is mathematically equivalent
to fitting independent $H_{0}^{\text{Pan+}}$ and $H_{0}^{\text{DES}}$.

Accordingly, Eq. \eqref{eq:joint-L-def} translates to
\begin{equation}
\chi_{\text{Joint}}^{2}(H_{0}^{\text{Pan+}},H_{0}^{\text{DES}};\boldsymbol{\varTheta})=\chi_{\text{Pan+}}^{2}(H_{0}^{\text{Pan+}};\boldsymbol{\varTheta})+\chi_{\text{DES}}^{2}(H_{0}^{\text{DES}};\boldsymbol{\varTheta})
\end{equation}
The covariance matrix of the joint Pantheon+/DES likelihood is block-diagonal.
If inter-survey correlations become available, the joint likelihood
can be generalized by replacing the block-diagonal covariance with
the corresponding full covariance matrix.

The nuisance parameters $M^{\text{Pan+}}$ or $M^{\text{DES}}$ (or
the corresponding $H_{0}^{\text{Pan+}}$ and $H_{0}^{\text{DES}}$)
can be integrated out analytically. Therefore, the profiled likelihood
shown in Fig. \ref{fig:Contours} is mathematically equivalent to
the likelihood obtained by marginalizing over $M^{\text{Pan+}}$ and
$M^{\text{DES}}$.\vspace{-.5cm}

\section{\label{app:Redshift-cutoff}\ \  Model comparison as a function
of the Maximum redshift cutoff}

The preference for the logarithmic luminosity-distance relation over
flat $\Lambda$CDM for the  post-ABC SNe Ia data, reported in Table
\ref{tab:best-fit-params}, is not exclusive to the highest-redshift
SNe Ia. To demonstrate this, we repeat the analysis of Section \ref{sec:Direct-fit}
using progressively higher maximum-redshift cuts. As shown in Table
\ref{tab:zmax}, the logarithmic relation first becomes favored at
$z_{\text{max}}\simeq0.4$, where $\Delta\text{BIC}$ exceeds $10$,
and the preference strengthens as higher-redshift data are included.

\begin{table}[H]
\begin{centering}
\caption{Comparison between the logarithmic luminosity-distance relation and
flat $\Lambda$CDM for the post-ABC Pantheon+ and DES joint sample
as a function of the maximum redshift included. Positive values of
$\Delta\chi^{2}$, $\Delta$AIC and $\Delta$BIC favor the logarithmic
relation. The onset of preference occurs at $z_{\text{max}}\simeq0.4$.
\label{tab:zmax}}
\par\end{centering}
\centering{}\footnotesize
\begin{tabular}{ccccrccrccr} \hline\hline\rule{0pt}{2.6ex}
$z_{\rm max}$ & \ \ $N_{\rm SNe}$ & \ $\chi^2_{\rm Loga}$ & \ $\chi^2_{\Lambda{\rm CDM}}$ & \ \ \ $\Delta\chi^2$ \ \ & AIC$_{\rm Loga}$ & AIC$_{\Lambda{\rm CDM}}$ & $\Delta$AIC & \ \ BIC$_{\rm Loga}$ & BIC$_{\Lambda{\rm CDM}}$ & $\Delta$BIC \\[0.6ex]
\hline\rule{0pt}{2.6ex}
0.2 & \ \ 1193 & \ \ 1387.9 & \ \ 1382.7 & \ \ -5.2 \ \  & 1391.9 & 1388.7 & \ \ -3.2 & \ \ 1402.0 & 1403.9 & +1.9 \\[0.1ex]
\bf{0.4} & \ \ \bf{2078} & \ \ \bf{2112.6} & \ \ \bf{2116.7} &  \ \ \bf{+4.1} \ \  & \bf{2116.6} & \bf{2122.7} &  \bf{+6.1} & \ \ \bf{2127.9} & \bf{2139.6} & \bf{+11.7} \\
0.6 & \ \ 2840 & \ \ 2772.4 & \ \ 2780.9 & \ \ +8.5 \ \ & 2776.4 & 2786.9 & +10.5 & \ \ 2788.3 & 2804.8 & +16.5  \\[0.1ex]
0.8 & \ \ 3361 & \ \ 3222.6 & \ \ 3235.9 & \ \ +13.3 \ \ & 3226.6 & 3241.9 & +15.3 & \ \ 3238.8 & 3260.3 & +21.5 \\[0.1ex]
1.0 & \ \ 3479 & \ \ 3344.6 & \ \ 3359.2 & \ \ +14.6 \ \ & 3348.6 & 3365.2 & +16.6 & \ \ 3360.9 & 3383.7 & +22.8 \\[0.1ex]
2.3 & \ \ 3521 & \ \ 3384.9 & \ \ 3402.0 & \ \ +17.1 \ \ & 3388.9 & 3408.0 & +19.1 & \ \ 3401.3 & 3426.5 & +25.3 \\[0.4ex]
\hline\hline \end{tabular}
\end{table}

\section{\label{app:Binning-Fig-1}\ \  Binning procedure for Figure \ref{fig:Direct-fit}}

The binned data shown in Fig. \ref{fig:Direct-fit} are intended solely
for visualization and do not enter the fitting procedure. Accordingly,
the full covariance matrix is not used; however, the weighted average
$\left\langle z\right\rangle $ and $\left\langle \Delta\mu\right\rangle $
are computed using the inverse variances from the covariance matrix
diagonal. Tables \ref{tab:raw_pan_buckets} and \ref{tab:age-corrected-pan-buckets}
summarize the corresponding binning scheme and binned data.

\begin{table}[H]
\begin{centering}
\caption{Binned pre-ABC Pantheon+ distance-modulus offsets, $\Delta\mu$, relative
to the fiducial Einstein--de Sitter model ($H_{0,{\rm EdS}}=73.00$
km/s/Mpc). The flat $\Lambda$CDM prediction, $\mu_{\Lambda{\rm CDM}}$,
is evaluated using $H_{0,\Lambda{\rm CDM}}=72.84$ km/s/Mpc and $\Omega_{\Lambda}=0.638$.\\ \label{tab:raw_pan_buckets}}
\par\end{centering}
\begin{centering}
\footnotesize
\begin{tabular}{cccccccc} \hline\hline\rule{0pt}{3ex}
\small{$z$} bin & \ \small{${N_{\rm SNe}}$} & \ \small{$\langle z\rangle$} & \ \small{$\langle\Delta\mu\rangle$} & \ $\sigma_\mu$ & \ $\Delta\mu_{\Lambda{\rm CDM}}$ & $\frac{\langle\Delta\mu\rangle-\Delta\mu_{\Lambda{\rm CDM}}}{\sigma_\mu}$ \\[1ex] \hline\rule{0pt}{2.6ex}
$[0.0,0.1)$ & \ \ 741 & \ \ 0.033 & \ \ 0.048 & \ \ 0.006 & 0.038 & $+1.81$ \\[0.1ex] $[0.1,0.2)$ & \ \ 207 & \ \ 0.158 & \ \ 0.140 & \ \ 0.009 & 0.148 & $-0.95$ \\[0.1ex]
$[0.2,0.3)$ & \ \ 259 & \ \ 0.247 & \ \ 0.206 & \ \ 0.009 & 0.214 & $-0.97$ \\[0.1ex]
$[0.3,0.4)$ & \ \ 186 & \ \ 0.340 & \ \ 0.291 & \ \ 0.010 & 0.274 & $+1.66$ \\[0.1ex]
$[0.4,0.5)$ & \ \ 98 & \ \ 0.443 & \ \ 0.341 & \ \ 0.014 & 0.330 & $+0.78$ \\[0.1ex]
$[0.5,0.6)$ & \ \ 81 & \ \ 0.545 & \ \ 0.356 & \ \ 0.018 & 0.379 & $-1.29$ \\[0.1ex]
$[0.6,0.7)$ & \ \ 54 & \ \ 0.645 & \ \ 0.423 & \ \ 0.024 & 0.419 & $+0.15$ \\[0.1ex]
$[0.7,1.0)$ & \ \ 50 & \ \ 0.762 & \ \ 0.432 & \ \ 0.025 & 0.460 & $-1.11$ \\[0.1ex]
$[1.0,2.3)$ & \ \ 25 & \ \ 1.383 & \ \ 0.723 & \ \ 0.045 & 0.601 & $+2.71$ \\[0.4ex]
\hline\hline\\ \end{tabular}
\par\end{centering}
\begin{centering}
\caption{Binned post-ABC Pantheon+ distance-modulus offsets, $\Delta\hat{\mu}$,
relative to the fiducial Einstein--de Sitter model ($H_{0,{\rm EdS}}=73.00$
km/s/Mpc). The flat $\Lambda$CDM prediction, $\hat{\mu}_{\Lambda{\rm CDM}}$,
is evaluated using $\hat{H}_{0,\Lambda{\rm CDM}}=73.06$ km/s/Mpc
and $\hat{\Omega}_{\Lambda}=0.462$, while the logarithmic relation
prediction, $\hat{\mu}_{{\rm loga}}$, uses $\hat{H}_{0,{\rm loga}}=72.56$
km/s/Mpc.\\ \label{tab:age-corrected-pan-buckets}}
\par\end{centering}
\centering{}\footnotesize
\begin{tabular}{ccccccccc} \hline\hline\rule{0pt}{3ex}
\small{$z$} bin & \ \small{${N_{\rm SNe}}$} & \ \small{$\langle z\rangle$} & \ \small{$\langle\Delta\hat\mu\rangle$} & \ $\hat\sigma_\mu$ & \ $\Delta\hat\mu_{\Lambda{\rm CDM}}$ & $\frac{\langle\Delta\hat\mu\rangle-\Delta\hat\mu_{\Lambda{\rm CDM}}}{\hat\sigma_\mu}$ & \ $\Delta\hat\mu_{\rm loga}$ & $\frac{\langle\Delta\hat\mu\rangle-\Delta\hat\mu_{\rm loga}}{\hat\sigma_\mu}$ \\[1ex]
\hline\rule{0pt}{2.6ex}
$[0.0,0.1)$ & \ \ 741 & \ \ 0.033 & \ \ 0.036 & \ \ 0.006 & \ \ 0.022 & $+2.40$ & \ 0.031 & $+0.88$ \\[0.1ex]
$[0.1,0.2)$ & \ \ 207 & \ \ 0.158 & \ \ 0.087 & \ \ 0.009 & \ \ 0.099 & $-1.43$ & \ 0.092 & $-0.64$ \\[0.1ex]
$[0.2,0.3)$ & \ \ 259 & \ \ 0.247 & \ \ 0.129 & \ \ 0.009 & \ \ 0.144 & $-1.64$ & \ 0.132 & $-0.28$ \\[0.1ex]
$[0.3,0.4)$ & \ \ 186 & \ \ 0.340 & \ \ 0.194 & \ \ 0.010 & \ \ 0.183 & $+1.14$ & \ 0.170 & $+2.40$ \\[0.1ex]
$[0.4,0.5)$ & \ \ 98 & \ \ 0.443 & \ \ 0.228 & \ \ 0.014 & \ \ 0.219 & $+0.61$ & \ 0.209 & $+1.29$ \\[0.1ex]
$[0.5,0.6)$ & \ \ 81 & \ \ 0.545 & \ \ 0.228 & \ \ 0.018 & \ \ 0.249 & $-1.20$ & \ 0.245 & $-0.98$ \\[0.1ex]
$[0.6,0.7)$ & \ \ 54 & \ \ 0.645 & \ \ 0.284 & \ \ 0.024 & \ \ 0.274 & $+0.41$ & \ 0.278 & $+0.26$ \\[0.1ex]
$[0.7,1.0)$ & \ \ 50 & \ \ 0.762 & \ \ 0.284 & \ \ 0.025 & \ \ 0.299 & $-0.61$ & \ 0.313 & $-1.19$ \\[0.1ex]
$\bf{[1.0,2.3)}$ & \ \ \bf{25} & \ \ \bf{1.383} & \ \ \bf{0.550} & \ \ \bf{0.045} & \ \ \bf{0.382} & $\bf{+3.74}$ & \ \bf{0.468} & $\bf{+1.83}$ \\[0.6ex]
\hline\hline \end{tabular}
\end{table}

\section{\label{app:Mu2}\ \  Diagnostic quadratic extension of the logarithmic
$d_{L}$ relation}

To assess possible deviations from the logarithmic luminosity-distance
relation \eqref{eq:dL-loga}, we consider its extension
\begin{equation}
d_{L}=\frac{c}{H_{0}}(1+z)\,\ln(1+z)\left[1+\delta\,\ln(1+z)+\mathcal{O}\left(\left(\ln(1+z)\right)^{2}\right)\right]\label{eq:Mu2-def}
\end{equation}
where $\delta$ is a dimensionless parameter characterizing the leading
quadratic correction to the logarithmic relation \eqref{eq:dL-loga}.
As reported in Table \ref{tab:kernel_mu2}, fitting Eq. \eqref{eq:Mu2-def}
to the joint Pantheon+ and DES sample yields $\delta=0.017\pm0.010$
for the  post-ABC data, consistent with zero at the $1.8\sigma$ level.
The Python scripts used to produce the quadratic logarithmic analysis
are provided at \small{\url{https://github.com/HoangNguyenUBB/SNeIa-loga-relation}}

\begin{table}[H]
\begin{centering}
\caption{Joint Pantheon+ and DES fits using the pure logarithmic relation ($\delta=0$)
and its quadratic extension \eqref{eq:Mu2-def}, with independent
$H_{0}$ values for the two surveys. The parameter $\delta$ is dimensionless.
The flat $\Lambda$CDM is shown as baseline. \\  \label{tab:kernel_mu2}}
\par\end{centering}
\centering{}\footnotesize \begin{tabular}{cccccccc} \hline\hline\rule{0pt}{2.4ex} & \multicolumn{3}{c}{Pre-ABC data} &\ \ & \multicolumn{3}{c}{Post-ABC data} \\[0.1ex]
\cline{2-4}\cline{6-8}\rule{0pt}{2.4ex} Parameter & Logarithmic & Quadratic & Flat $\Lambda$CDM &\ \ & Logarithmic & Quadratic & Flat $\Lambda$CDM \\[.4ex]
\hline\rule{0pt}{2.6ex}
$H_0^{\mathrm{Pan+}}$\! (km/s/Mpc) & $70.64\pm0.12$ & $72.42\pm0.18$ & $73.03\pm0.18$ &\ \ & $72.56\pm0.12$ & $72.78\pm0.18$ & $73.34\pm0.18$ \\[0.1ex]
$H_0^{\mathrm{DES}}$ (km/s/Mpc) & $65.66\pm0.14$ & $68.80\pm0.28$ & $69.50\pm0.27$ &\ \ & $68.79\pm0.15$ & $69.20\pm0.27$ & $69.79\pm0.29$ \\[0.1ex]
$\small\pmb{\delta}$ & --- & $\boldsymbol{0.140\pm0.010}$ & --- &\ \ & --- & $\boldsymbol{0.017\pm0.010}$ & --- \\[0.1ex]
$\small\pmb{\delta/\sigma_\delta}$ & --- & $\boldsymbol{13.52}$ & --- &\ \ & --- & $\boldsymbol{1.78}$ & --- \\[0.1ex]
$\Omega_\Lambda$ & --- & --- & $0.656\pm0.012$ &\ \ & --- & --- & $0.491\pm0.014$ \\[0.1ex]
$\chi^2_{\min}$ & 3583.6 & 3386.5 & 3386.1 &\ \ & 3384.9 & 3381.7 & 3402.0 \\[0.1ex]
AIC & 3587.6 & 3392.5 & 3392.1 &\ \ & 3388.9 & 3387.7 & 3408.0 \\[0.1ex]
BIC & 3599.9 & 3411.0 & 3410.6 &\ \ & 3401.3 & 3406.2 & 3426.5 \\[0.4ex]
\hline\hline \end{tabular}
\end{table}

\section{\label{app:Kernel-forms}\ \  Luminosity-distance Kernel formulation}

For any spatially flat Friedmann--Lema\^itre--Robertson--Walker
(FLRW) metric, the luminosity distance can be written as $d_{L}=(1+z)\int_{a}^{1}\frac{c}{a'H}d\ln a'$,
where $a=(1+z)^{-1}$. Differentiation with respect to $\ln(1+z)$
gives
\begin{equation}
\frac{H_{0}}{c}\frac{d\left[d_{L}/(1+z)\right]}{d\ln(1+z)}=\frac{H_{0}}{aH(a)}\equiv K(z)
\end{equation}
which defines a \emph{dimensionless} quantity, which we refer to as
the luminosity-distance kernel (or kernel for brevity). The kernel
$K$ can be determined \emph{empirically} from the observables $z$
and $d_{L}$ through $\frac{H_{0}}{c}\frac{d\left[d_{L}/(1+z)\right]}{d\ln(1+z)}$,
while each cosmological model predicts a corresponding theoretical
form. In particular, the logarithmic relation \eqref{eq:dL-loga}
predicts that $K=1$, whereas a flat $w$CDM cosmology gives $K_{w\text{CDM}}=\frac{H_{0}}{aH_{w\text{CDM}}(a)}=\frac{1+z}{\sqrt{(1-\Omega_{\text{DE}})(1+z)^{3}+\Omega_{\text{DE}}(1+z)^{3(1+w)}}}$.
For $w=-1$, $K_{w\text{CDM}}$ reduces to the flat $\Lambda$CDM
kernel used to plot $\Lambda$CDM kernel curves in Fig. \ref{fig:K-bucket}
of Section \ref{sec: Kernel-bucket}.

For the observational analysis, the kernel is estimated as \emph{the
slope of a weighted linear regression} of $y\equiv\frac{H_{0}}{c}\frac{d_{L}}{1+z_{\text{hel}}}$
against $u\equiv\ln(1+z_{\text{HD}})$ within each redshift interval
$z_{\text{HD}}\in[z_{\text{lo}},z_{\text{hi}})$, using inverse variances
from the covariance matrix diagonal as weights.

Implementation details for the observational kernel estimation are
given in Appendix \ref{app:Binning-Fig-3}.

\section{\label{app:Binning-Fig-3}\ \  Binning procedure for Figure \ref{fig:K-bucket}}

The observational kernel estimation in Section \ref{sec: Kernel-bucket}
proceeds as follows:
\begin{itemize}
\item For the pre-ABC data, each dataset (Pantheon+ and DES) is first fitted
separately with flat $\Lambda$CDM to obtain $H_{0}^{\text{Pan+}}$
and $H_{0}^{\text{DES}}$. These values, $H_{0}^{\text{Pan+}}=73.03$
and $H_{0}^{\text{DES}}=69.50$ km/s/Mpc, are then used to construct
the corresponding $y$ variables before performing the linear regression
against $u$.
\item For the  post-ABC data, the same procedure is applied, except that
$H_{0}^{\text{Pan+}}=72.56$ and $H_{0}^{\text{DES}}=68.79$ km/s/Mpc
are obtained from fitting the logarithmic luminosity-distance relation
\eqref{eq:dL-loga}.
\end{itemize}
Tables \ref{tab:K-bucket-raw} and \ref{tab:K-bucket-age-corrected}
list the kernel estimates for the pre-ABC and  post-ABC samples, respectively.
The first five bins are non-overlapping and are\emph{ chosen to yield
approximately comparable statistical uncertainties}. The remaining
five bins are interlaced with the first set.

\begin{table}[H]
\begin{centering}
\caption{Kernel estimates for the ten redshift buckets used in Fig. \ref{fig:K-bucket}
for the pre-ABC Pantheon+ and DES joint sample. The last two columns
give the deviations from the logarithmic prediction ($K=1$) and the
best-fit flat $\Lambda$CDM theoretical prediction ($\Omega_{\Lambda}=0.656$),
respectively. \\ \label{tab:K-bucket-raw}}
\par\end{centering}
\begin{centering}
\footnotesize
\begin{tabular}{cccrcccccc} \hline\hline\rule{0pt}{3ex}
Bin & \normalsize{$z_{\rm lo}$} & \normalsize{$z_{\rm hi}$} & \small{$N_{\rm SNe}$} & \normalsize{$\langle z\rangle$} & \small{$K_{\rm fit}$} & \small{$\sigma_K$} & \small{$K_{\Lambda{\rm CDM}}$} & \normalsize{$\frac{K_{\rm fit}-1}{\sigma_K}$} & \normalsize{$\frac{K_{\rm fit}-K_{\Lambda{\rm CDM}}}{\sigma_K}$} \\[1ex] \hline\rule{0pt}{2.4ex}
1  & 0.000 & \ 0.010 & \ \ 111  & \ 0.007 & \ 1.025 & \ 0.008 & \ 1.003 & \ \ $+3.15$ & $\mathbf{+2.75}$ \\[0.1ex]
2  & 0.010 & \ 0.035 & \ \ 513  & \ 0.024 & \ 0.998 & \ 0.011 & \ 1.011 & \ \ $-0.19$ & $\mathbf{-1.25}$ \\[0.1ex]
3  & 0.035 & \ 0.150 & \ \ 420  & \ 0.072 & \ 1.023 & \ 0.008 & \ 1.032 & \ \ $+2.83$ & $\mathbf{-1.02}$ \\[0.1ex]
4  & 0.150 & \ 0.500 & \ 1405 & \ 0.327 & \ 1.098 & \ 0.007 & 1.098 & $+14.21$ & $\mathbf{-0.11}$ \\[0.1ex]
5  & 0.500 & \ 2.300 & \ 1072 & \ 0.676 & \ 1.121 & \ 0.020 & \ 1.111 & \ \ $+6.09$ & $\mathbf{+0.48}$ \\[0.4ex]
\hline\rule{0pt}{2.4ex}
6  & 0.008 & \ 0.027 & \ \ 354  & \ 0.018 & \ 0.991 & \ 0.012 & \ 1.009 & \ \ $-0.77$ & $\mathbf{-1.48}$ \\[0.1ex]
7  & 0.027 & \ 0.100 & \ \ 515  & \ 0.045 & \ 1.020 & \ 0.009 & 1.021 & \ \ $+2.13$ & $\mathbf{-0.03}$ \\[0.1ex]
8  & 0.100 & \ 0.270 & \ \ 542  & \ 0.200 & \ 1.051 & \ 0.011 & \ 1.073 & \ \ $+4.47$ & $\mathbf{-1.99}$ \\[0.1ex]
9  & 0.270 & \ 0.700 & \ 1705 & \ 0.472 & \ 1.100 & \ 0.008 & \ 1.112 & $+11.73$ & $\mathbf{-1.42}$ \\[0.1ex]
10 & 0.700 & \ 2.300 & \ \ 332  & \ 0.856 & \ 1.089 & \ 0.049 & \ 1.098 & \ \ $+1.82$ & $\mathbf{-0.19}$ \\[0.4ex]
\hline\rule{0pt}{2.6ex}
Average & & & & & & & & \ \ $+4.55$ & $\mathbf{-0.43}$ \\[0.6ex]
RMS     & & & & & & & & \ \ \ \ \ $6.50$ & \ \ $\mathbf{1.36}$ \\[0.4ex]
\hline\hline\\ \end{tabular}\caption{Kernel estimates for the ten redshift buckets used in Fig. \ref{fig:K-bucket}
for the post-ABC Pantheon+ and DES joint sample. The last two columns
give the deviations from the logarithmic prediction ($K=1$) and the
best-fit flat $\Lambda$CDM theoretical prediction ($\hat{\Omega}_{\Lambda}=0.491$),
respectively. \\ \label{tab:K-bucket-age-corrected}}
\par\end{centering}
\centering{}\footnotesize
\begin{tabular}{cccrcccccc} \hline\hline\rule{0pt}{3ex}
Bin & \normalsize{$z_{\rm lo}$} & \normalsize{$z_{\rm hi}$} & \small{$N_{\rm SNe}$} & \normalsize{$\langle z\rangle$} & \small{$\hat K_{\rm fit}$} & \small{$\hat\sigma_K$} & \small{$\hat K_{\Lambda{\rm CDM}}$} & \normalsize{$\frac{\hat K_{\rm fit}-1}{\hat\sigma_K}$} & \normalsize{$\frac{\hat K_{\rm fit}-\hat K_{\Lambda{\rm CDM}}}{\hat\sigma_K}$} \\[1ex]
\hline\rule{0pt}{2.4ex}
1  & 0.000 & \ 0.010 & \ \ 111  & \ 0.007 & \ 1.017 & \ 0.008 & \ 1.002 & \ \ $\mathbf{+2.19}$ & $+1.98$ \\[0.1ex]
2  & 0.010 & \ 0.035 & \ \ 513  & \ 0.024 & \ 0.981 & \ 0.010 & \ 1.005 & \ \ $\mathbf{-1.78}$ & $-2.29$ \\[0.1ex]
3  & 0.035 & \ 0.150 & \ \ 420  & \ 0.072 & \ 0.991 & \ 0.008 & \ 1.014 & \ \ $\mathbf{-1.13}$ & $-2.86$ \\[0.1ex]
4  & 0.150 & \ 0.500 & \ 1405 & \ 0.327 & \ 1.014 & \ 0.007 & \ 1.024 & \ \ $\mathbf{+2.09}$ & $-1.54$ \\[0.1ex]
5  & 0.500 & \ 2.300 & \ 1072 & \ 0.676 & \ 1.004 & \ 0.018 & \ 0.986 & \ \ $\mathbf{+0.24}$ & $+0.98$ \\[0.4ex]
\hline\rule{0pt}{2.4ex}
6  & 0.008 & \ 0.027 & \ \ 354  & \ 0.018 & \ 0.978 & \ 0.012 & \ 1.004 & \ \ $\mathbf{-1.88}$ & $-2.23$ \\[0.1ex]
7  & 0.027 & \ 0.100 & \ \ 515  & \ 0.045 & \ 0.995 & \ 0.009 & \ 1.009 & \ \ $\mathbf{-0.55}$ & $-1.56$ \\[0.1ex]
8  & 0.100 & \ 0.270 & \ \ 542  & \ 0.200 & \ 0.990 & \ 0.011 & \ 1.025 & \ \ $\mathbf{-0.89}$ & $-3.18$ \\[0.1ex]
9  & 0.270 & \ 0.700 & \ 1705 & \ 0.472 & \ 0.997 & \ 0.008 & \ 1.012 & \ \ $\mathbf{-0.43}$ & $-1.98$ \\[0.1ex]
10 & 0.700 & \ 2.300 & \ \ 332  & \ 0.856 & \ 0.975 & \ 0.045 & \ 0.959 & \ \ $\mathbf{-0.56}$ & $+0.36$ \\[0.4ex]
\hline\rule{0pt}{2.6ex}
Average & & & & & & & & \ \ $\mathbf{-0.27}$ & $-1.23$ \\[0.6ex]
RMS     & & & & & & & & \ \ \ \ \,$\mathbf{1.37}$  & \ \ \,$2.05$ \\[0.4ex]
\hline\hline \end{tabular}
\end{table}

\end{document}